\documentclass{article} 
\usepackage{graphicx}
\usepackage{subcaption}
\usepackage{nips14submit_e,times}
\usepackage{url}
\usepackage[english]{babel}
\usepackage[T1]{fontenc}
\usepackage[utf8]{inputenc}
\usepackage{listings}
\usepackage{float}
\usepackage{subfiles}
\providecommand{\keywords}[1]{\textbf{\textit{Keywords---}} #1}
\usepackage{booktabs} 
\usepackage{multirow}
\usepackage[export]{adjustbox}[2011/08/13]

\title{Technology-driven Alteration of Nonverbal Cues and its Effects on Negotiation}

\author{
Raiyan Abdul Baten\textsuperscript{1}, Ehsan Hoque\textsuperscript{2,*}\\
\textsuperscript{1}Department of Electrical and Computer Engineering, University of Rochester, NY, USA\\ \textsuperscript{2}Department of Computer Science, University of Rochester, NY, USA\\
\texttt{*mehoque@cs.rochester.edu}
}

\nipsfinalcopy 

\begin{document}

\maketitle
\begin{abstract}
A person’s appearance, identity, and other nonverbal cues can substantially influence how one is perceived by a negotiation counterpart, potentially impacting the outcome of the negotiation. With recent advances in technology, it is now possible to alter such cues through real-time video communication. In many cases, a person’s physical presence can explicitly be replaced by 2D/3D representations in live interactive media. In other cases, technologies such as deepfake can subtly and implicitly alter many nonverbal cues—including a person’s appearance and identity—in real-time. In this article, we look at some state-of-the-art technological advances that can enable such explicit and implicit alteration of nonverbal cues. We also discuss the implications of such technology for the negotiation landscape and highlight ethical considerations that warrant deep, ongoing attention from stakeholders.
\end{abstract}


\section*{}
In the recent movie \textit{The Irishman}, filmmakers used artificial intelligence (AI) to make the actors—Al Pacino, Robert De Niro, and Joe Pesci, all in their seventies—look decades younger. Although the actors talked and conducted themselves just as they do in the present day, the technology enabled (or deluded) viewers to see their characters in a much earlier time. This same technology could enable you to become a different self the next time you negotiate remotely. You may not choose to use such technology but should you worry that your counterparts will? In some ways, it is already happening.

Negotiation is at its core a form of social interaction—one that is shaped heavily by how each party and its actions are perceived by the other. Your negotiation counterparts are your audience, just as you are theirs. Being open to listening more, empathizing when appropriate, and taking another’s perspective are essential elements of negotiation, which are often guided by a careful observation of the verbal and non-verbal cues of the negotiation counterparts. Importantly, a person’s aural, visual, and other nonverbal cues play a crucial role in shaping the perceptions of others, and it is suggested that up to 93\% of the meaning of a message can be conveyed by facial and vocal cues rather than speech~\cite{mehrabian1971silent}. Individuals who are good at sensing and adapting to these dimensions are viewed as good negotiators or mediators. Imagine if the technology exists to allow individuals to dynamically change some of these non-verbal attributes to their advantage during negotiations. What if people could modify their race, gender, age, speaking style, eye contact, or even expressions during a video call? What are the circumstances in which this practice may be viewed as ethical? What would make it unethical? If such technology becomes widely available, how can we ensure that it is used in a fair, unbiased, and transparent way during a negotiation? 

Such possibilities may seem far-fetched to many. However, with recent advances in online and AI-driven communication technology, it is possible today to alter people’s appearances, identities, and other nonverbal cues on the fly. This has opened up myriad new opportunities as well as risks that have the potential to impact the future of negotiations—opportunities and risks that have gone under-addressed in the negotiation literature. In this article, we highlight some state-of-the-art technologies that can alter appearances and perceptions, discuss the consequences and ethical considerations attendant on the use of such technologies, and consider futuristic possibilities.

\section*{Online Interactions and Negotiations}

The use of remote communication tools for day-to-day business interactions has increased greatly due to the lockdowns and social distancing measures brought on by the COVID-19 pandemic. Video conferencing services like Zoom and Microsoft Teams, and team collaboration services such as Slack, have become the go-to work-from-home tools for businesses around the globe. Ongoing research efforts—for example, the development of virtual office technologies using augmented reality and virtual reality (AR/VR)—have been accelerated to better simulate real-life interactions online.

While most of these tools attempt to bridge the gap between face-to-face and online communications, they can widen the gap as well. Many remote communication platforms enable people to alter explicitly or implicitly their appearance and other nonverbal cues in real-time—something not possible in face-to-face interactions. For example, 2D/3D avatars that depict humans in virtual interaction environments can explicitly replace a person’s unique natural presence with a standard graphical one. Again, technologies such as deepfake can implicitly alter the perceptions of one’s race, speaking style, age, and other attributes, often in real-time~\cite{suwajanakorn2017synthesizing}. To comprehend fully the importance of such nonverbal cues, consider a study by Bertrand et al.~\cite{bertrand2004emily}. The researchers sent out résumés for job applications; the résumés were identical except for variation in names, which were chosen to suggest the “applicant’s” race. The employers’ responses to the résumés varied significantly depending on the perceived race of the applicant. If names alone have such an effect, it is of no surprise that seeing and hearing someone in a modified way has all the more power to affect human interactions. 

We explore technology-driven explicit and implicit alteration possibilities in the next two sections.

\section*{User Interface Scaffolding: Explicit Alterations in Appearances and Identities}

While limited efforts have been made in exploring user interface scaffolding in the domain of negotiation, the opportunities and constraints therein have nonetheless been explored in other closely related interaction contexts, such as online collaboration among team members or peer-interactions in online social networks. We document two example explorations below.

Consider online collaboration within a team. The increased prevalence of working from home has curtailed many of the benefits of face-to-face interactions found in traditional workplaces. However, working from home also has opened up opportunities that can potentially improve how teams work. Whiting et al.~\cite{whiting2020parallel,whiting2020can} temporarily masked the identities of team members by using pseudonyms in a real-time online chatroom messaging platform. By using two-way pseudonym masking, the researchers created “parallel worlds” where team members believed they were collaborating with different people although they actually were communicating with the same set of teammates many times over. This resulted in enhanced team viability—a team’s capacity to sustain collaboration—by affording people fresh starts with renewed identities. 

In another study, Baten et al.~\cite{baten2020availability} controlled the exposure of demographic cues (gender and race) by using 2D avatars and examined the effects on creative outcomes. The participants sought creative inspiration from their peers in an online social network, across multiple rounds of creative idea-generation tasks. They submitted their ideas in text format, which could be viewed by their followers. The researchers observed that when the avatars were shown, the same-gender connections became significantly stable with time—in other words, if someone followed a peer of his/her own gender for creative inspiration, they were likely to keep following that person in future rounds. This behavior was not exhibited in the demography-agnostic control group. The researchers also found that creative ideas tend to be significantly more homogeneous within demographic groups than between. This implies that choosing to follow peers based on self-similarity can hurt the diversity of ideas one can take inspiration from. In the end, the mere exposure of gender and race cues corresponded to a significantly reduced variety in the peer-stimulated creative ideas—a rather counterproductive outcome.

Using self-reported surveys (to evaluate team viability) and natural language processing (to assess creativity), both studies—Whiting’s study of pseudonymous altered identities and Baten’s study of masked identities—illustrate how user interface scaffolding can influence important aspects of human performance. To understand the idea of user interface scaffolding as it relates to negotiation, let us take a step back and consider the differences among video conferences, telephone calls, and text messages. The three media provide a decreasing level of communication richness—telephone calls and text messages lack visual cues, and text messages lack audio cues as well. Parties interacting over the various media are aware of the missing cues, and this knowledge can affect their sense of depersonalization and anonymity and influence their communication strategies~\cite{purdy2000impact}. Both Whiting and Baten studied interactions that lacked audio-visual cues. However, in stark contrast to traditional text messaging, these two studies attempted to \textit{alter} or \textit{mask} the true identities of the interacting parties by using pseudonyms and avatars and were able to influence the interaction outcomes substantially. Such alteration or masking are feats easily achieved in online communication and can lead to myriad possibilities in negotiations. What if certain bargaining is not going in your favor and you had the opportunity to make a fresh start with a different identity? What if demographic cues were completely removed, or at least standardized, in order to break stereotypical notions that can give either party an advantage over the other? Female gamers often mask their identities and adopt fake male avatars to avoid perceptions of diminished competence in online gaming communities~\cite{kaye2017exploring}, illustrating a use-case of such renewed identity. However, it remains less clear how negotiating parties would adapt their strategies to such masking possibilities in online negotiations. 

User interface scaffolding as illustrated above is no longer a futuristic possibility, and viable products that incorporate such capabilities are already hitting the market. With video conferencing technologies widely adopted, services like the Snap Camera~\cite{snap2020camera} are being introduced. These services allow a user to make real-time alterations of audio-visual cues. People can overlay their video streams with artifacts of their choice, potentially altering how their counterparts perceive them and their actions. 

\begin{figure}
    \centering
    \includegraphics[width=1\linewidth]{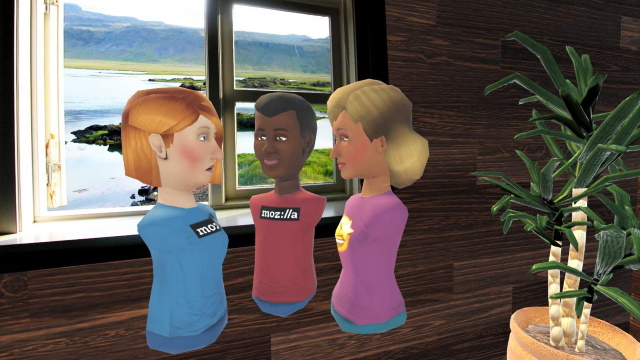}
    \caption{“Hubs by Mozilla” allows people to join in collaborative virtual environments using their browsers. The participants can choose their own avatars and use voice and text to communicate naturally with their peers. Image is courtesy of Mozilla.}
    \label{moz}
\end{figure}

In contrast with video conferencing, where the interacting parties are constrained to a box on a screen, virtual reality-based spaces offer a more immersive experience~\cite{ruvimova2020transport}. For instance, mobile knowledge workers can interact with each other in virtual offices, using AR/VR headsets~\cite{fereydooni2020virtual,ofek2020towards} or just their browsers~\cite{hubs2020mozilla} (see Figure~\ref{moz}). Such virtual interactions are typically carried out via 3D agents that represent their respective users. The users’ gestures, facial expressions, eye contact, body movements, and physical appearances become completely altered, and all of the parties need to adjust their interaction strategies accordingly.

Consider the example of buying a car with the parties interacting face-to-face. In real life, this is a very stressful process for the buyer and a very high-stakes encounter for the salesperson, who may have daily quotas to meet. A buyer wants to obtain objective information in order to make a financially sound purchase, whereas the seller wants to develop a relationship and a sense of rapport that will make it difficult for the buyer to walk away from the interaction. Experienced salespeople of cars and other goods often classify “qualified” prospective buyers on the basis of their clothes, speaking style, and other nonverbal cues. Conducting the negotiation in a virtual showroom—where both parties are represented by 3D avatars that alter their appearances and other nonverbal cues—can potentially impact how the interaction unfolds. In many cases, the virtual setting can be more advantageous for buyers; it can allow them to end encounters with salespeople without any social consequences. On the other hand, a salesperson’s ability to identify prospective buyers and develop trusting relationships with them can be disrupted in a virtual setting. This will necessitate a review of the salesperson’s strategy, and call for improvisations on how such technology can be used to the sellers’ advantage.

Although holding a conversation using 3D agents in a virtual environment may appear unnatural, such settings allow people to have difficult conversations without the stress of being physically colocated. The dilemma of whether to negotiate on the counterpart’s turf or to insist that the negotiation take place on one’s own literally vanishes, as a shared virtual space emerges as a neutral venue. To further underscore this phenomenon, let us review a historic negotiation from 1986 between US President Ronald Reagan and General Secretary Mikhail Gorbachev of the former USSR. They agreed to negotiate to end the Cold War and to hold their meeting in Reykjavik, Iceland, which was equidistant from the United States and the Soviet Union and viewed as neutral ground~\cite{reynolds2016speaking}. On the day of the negotiation, Reagan purposefully entered the venue earlier than Gorbachev. When Gorbachev arrived, Reagan stepped out the door as though welcoming Gorbachev into his home~\cite{trip1986}. Many viewed this gesture as a power play to show dominance even on neutral ground. Technology, particularly virtual reality-based environments, eliminates such moves and allows for enhanced neutralization of power between the parties.   

These developments pose unique questions and concerns: 
\begin{itemize}
    \item How would you feel if your counterpart used a real-time video filter to appear more likable and convincing? 
    \item If you are a person who makes good use of your naturally charismatic flair in face-to-face interactions, how would you adapt your negotiation strategies if you were represented by a standardized 3D agent? 
    \item What happens if either party feels threatened by the appearance and other nonverbal cues of a counterpart’s avatar? 
    \item How do these technologies impact human relationships and negotiation outcomes? 
\end{itemize}

In the conventional negotiation landscape, job applicants trying to put on their best face in interviews and car salespeople trying to seem friendly are common scenarios. Such parties will increasingly and inevitably use new technologies to help achieve their respective goals. It is thus crucial that the promises and drawbacks of such technologies are carefully considered.

\section*{Deepfakes: Covert Changes in Audio-Visual Cues}

The cautionary message of “You can’t believe everything you see on the Internet” was strikingly illustrated by actor Jordan Peele’s viral impersonation of President Barack Obama~\cite{BuzzFeedVideo18}. In the video, AI-constructed footage of Obama convincingly appeared to show him delivering a speech that was, in fact, delivered by Peele. This technology, popularly known as “deepfake,” has invited a lot of controversy in recent times, as fabricated audio-visual products have started to bias public discourse by spreading falsehoods. Deepfakes—images or recordings that have been convincingly altered and manipulated to misrepresent someone as doing or saying something that was not actually said or done—achieve this by superimposing a target person’s identifying features on top of a content template, so that the target person is seen delivering the desired, often malicious, content~\cite{suwajanakorn2017synthesizing}. The use of such technologies extends to online and real-time communications as well, holding strong implications for the field of negotiation.

Imagine you are negotiating with a person who you know will view you more favorably if they thought you were from a certain demographic group in terms of nationality, race, or gender, for instance. An exemplary case can be drawn from the information and communications technology (ICT) industry, where Indian call center agents pose as Americans when working for US-outsourced firms~\cite{poster2007s}. Adopting the American accent, speaking style, and grammar become part of the skill set required for agents so that they can deal most effectively with clients. One can further imagine scenarios where a person’s racial cues, such as skin tone and accent, can dictate the quality and outcome of a given interaction if the other party is likely to be influenced by a counterpart’s race. Aging upward or downward, or changing the perception of one’s gender, can similarly allow people to influence how they are received by their counterparts. Using deepfakes, it is possible to change a person’s identifying attributes such as skin tone, hair color, gender~\cite{lu2018attribute}, age~\cite{antipov2017face}, accent, and speech pattern~\cite{sisman2020overview}. Many of these deepfakes can already be generated on the fly, and it is only a matter of time before all such conversions are possible in real-time. Thus, a call center agent in the above example can simply speak naturally while the technology converts her accent in real-time. This technology can make life easy for people across a wide spectrum of human activity. On the downside, using deepfake technology in this way remains questionable from an ethical perspective, especially when there is a lack of transparency about its use to all the interacting parties.

As another example, consider Parkinson’s disease patients, who predominantly suffer from tremors in body movements and vocal sounds. Such tremors can be perceived by many as signaling a lack of competence and influence. Deepfake technologies can allow a quick fix to this problem by smoothing out the gestures and vocal sounds in real-time or in post-processing of video contents. Such alterations need not be covert; they can be revealed to all the interacting parties in advance. Yet in the context of a negotiation, these alterations can induce feelings of power and increased confidence in someone with Parkinson’s or a similar disease, potentially altering the course of the interaction itself.

Modifying the self-perception of one’s \textit{own} behavior in real-time also has implications for negotiation. In their research, Costa et al.~\cite{costa2018regulating} recognized that emotions play a crucial role in how interpersonal conflicts unfold. They artificially altered people’s own voice feedback as they communicated with a partner via Skype (audio-only conversations). In other words, if Seller A was talking to Buyer B in this setting, Seller A would hear her own voice in an artificially altered manner in real-time, while Buyer B received the unaltered audio. The results showed positive outcomes: automatically making the self-perception of one’s voice calmer than it really is makes one less anxious, while a lowered pitch during a contentious debate makes a person feel more powerful. On the other hand, there can be potential drawbacks to such technology. This particular study did not alter the audio cues as heard by the receiving party in the interaction, but doing so is, as previously noted, technologically feasible, raising ethical concerns around agency and consent.

\begin{figure}
    \centering
    \includegraphics[width=1\linewidth]{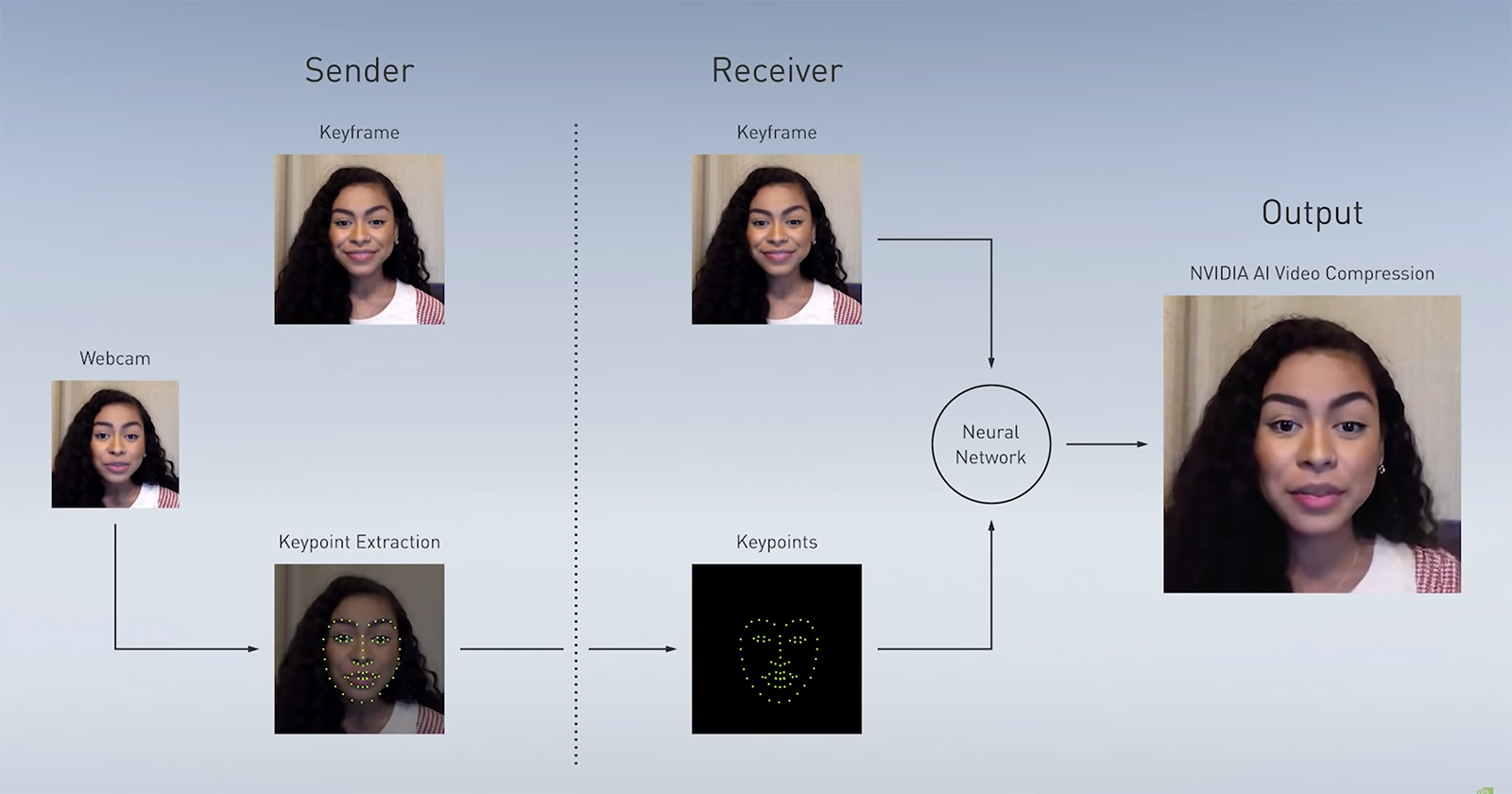}
    \caption{NVIDIA’s video compression technology uses AI to deliver high-quality output over slow Internet connections. A single reference image of the sender is first sent to the receiver. The temporal locations of various facial landmarks are later sent, from which the output is reconstructed at the receiving end. While the technology is making rapid progress, ``uncanny valleys'' can still take place at the receiving end due to imperfect reconstruction. Image courtesy is of NVIDIA.}
    \label{nvidia}
\end{figure}

The indirect and unintended alteration of audio-visual cues is a potential side effect of AI-powered communication media, where solving one technical challenge can result in introducing a different one. Take, for example, NVIDIA’s recent video compression technology, which attempts to make real-time video communication seamless even over slow Internet connections~\cite{Schneider20} (see Figure~\ref{nvidia}). The technology first sends a reference image (“keyframe”) of the caller to the receiver. Then, for each video frame, it sends only the locations of key facial landmarks around the eyes, nose, and mouth, so that the full video can be reconstructed at the receiver’s end by dynamically moving the respective facial landmarks of the keyframe image in real-time. Such technology can substantially reduce the required bandwidth for video transmission, but it can also introduce distortion in the visual cues at the receiving end. The reconstructed images/videos often fall within the “uncanny valley,” where they appear very close to the true references but are not \textit{exact} depictions, which can be unsettling to the human eye~\cite{broad2020amplifying}. In many cases, however, the uncanny valley can help achieve the desired goal. If you were to negotiate over video conferencing and the technology improved your video quality substantially at the cost of putting your facial cues within the uncanny valley, would you accept the trade-off? How would this affect the building of trust and a lasting relationship with your counterpart, beyond achieving utilitarian goals? These questions require serious and careful thought as the technology matures over time.

In all the examples of deepfake technology discussed above, audio-visual cues are modified in online communication media, often unbeknownst to either or both of the interacting parties—something rarely ever possible in face-to-face interactions. If negotiation is challenging when audio-visual cues are missing—for example, over the telephone or in text messaging—the situation is made substantially more complicated by the introduction of \textit{covertly altered} cues, a novel reality in the negotiation landscape.

\section*{Ethical Considerations}
In user interface scaffolding, explicit alterations in identities are generally known to both of the interacting parties, whereas in deepfakes, the alterations can potentially be hidden from either or both. This raises several ethical concerns, including: To what extent should technology be allowed to modify the parties’ interactions, and who should have the agency to decide?

It is important to note that some of the implications described in this article are unique to negotiations that are technology powered—for example, those resulting from real-time alteration of racial signifiers such as skin color and accent. However, some of the technology-driven alterations are supercharged versions of events that come up often in conventional negotiation environments, such as attempts to appear more presentable in job interviews with the help of video filters. Given that the technology is still in its infancy, how negotiators in various contexts will perceive such technology-driven modifications is not yet clear. Notifying all of the interacting parties fully about the implications of the technologies in use and obtaining informed consent prior to the interaction can potentially bring an acceptable level of transparency to the negotiation setting. However, we acknowledge there may be circumstances when nondisclosure of automatic cue-alterations can be justified—for example, in life-threatening cases such as hostage situations.

Beyond the concerns around agency and informed consent, a broader ethical question looms: Should people be \textit{allowed} to change their age, gender, or race cues, knowing that such actions can alter the course of the negotiation? One can question the morality of allowing people to do so, citing the ethical drawbacks of deceptive behavior. On the other hand, arguments can be made that discrimination against people based on race, gender, and age groups can be mitigated by using these technologies, to create a fairer level-playing field.

\section*{Recommendations}
Previous research has noted principles and guidelines for developing human-computer interaction systems, in particular, human-AI systems. For instance, Amershi et al.~\cite{amershi2019guidelines} discussed design guidelines for AI-infused systems, which have AI-harnessing features directly exposed to the end-user (such as e-commerce websites with product recommendations and social media news feeds). As we have discussed, in the context of negotiation, the technology-driven modifications in identities, appearances, and other nonverbal cues can be explicit or covert to the interacting parties. We build on previous work to make the following recommendations for developing negotiation-centric tech products:

\begin{enumerate}
    \item \textbf{Make clear upfront what the system can do.} To impose transparency, any technology-driven artifact or modification used during negotiations should always be declared upfront by the platform. This is particularly important in technologies that can implicitly or covertly change nonverbal cues in a negotiation setting. For example, if an AI-driven system is changing a person’s accent or filtering out his tremor, this should be made clear upfront to all the interacting parties. 
    
    \item \textbf{Make clear how well the system can do what it can do.} Managing expectations is crucial when much of the communication richness is handed over to the technology-powered interaction medium. For example, in the case of the video compression technology mentioned above, the developers should ensure fully informing users by clearly stating how well the technology can reconstruct facial features, warning users about uncanny valleys in the reconstructed video frames, and clarifying the technology’s potential usefulness and consequences.
    
    \item\textbf{Be mindful of social biases.} AI-driven systems should not reinforce undesirable and unfair social biases and stereotypes. In many cases, such biases can unintentionally get incorporated into intelligent systems if the corresponding training data contain stereotypical patterns of human communication. System designers should proactively attempt to collect diversified and fair sets of data and be mindful of potential biases in the interface and intervention designs.
\end{enumerate}

Using technology to mask and potentially upgrade a negotiator’s appearance and behavior is a reality that is rapidly taking shape. The overarching challenge is to determine where the boundaries are, who sets them, and what—if any—monitoring and enforcement options are possible. Answering these questions will warrant deep, ongoing consideration. Meanwhile, individuals should consider, as best they can, whether and how to use such technology. Organizations should develop provisional policies on its use. Just as people should be on their guard against AI-improvised materials on the Internet, negotiators should be aware of the fact that when interacting online, what they see and hear from others may have artificial elements.

\section*{Conclusion}
Reading the other party’s walkaway value, their apparent interest in making a deal, and their behavior often determines how a given negotiation unfolds. Unfortunately, even in face-to-face interactions—arguably the richest form of communication—such information is typically incomplete and subject to interpretation. The dynamic process of making offers and assessing counteroffers helps both sides to better determine the true position of the other~\cite{wheeler2002negotiation}. With the advent of modern remote communication technologies, explicit and implicit modifications of nonverbal cues have now become realities, introducing new dimensions of complexity into the dynamics of negotiation like never before. 

Negotiators fearful of exploitation and seeking to guard their interests may welcome the added layer of privacy that representation by a poker-faced avatar affords. On the other hand, a car salesperson honestly trying to give a customer the best deal possible and to communicate that sincerity may be frustrated at the diminished level of trust and respect that a virtual agent is able to command. Emotions, power differentials, and status are radically changed in these altered settings. Perception is a core component of negotiation dynamics, and computer-mediated alterations to identities, behaviors, and audio-visual cues can significantly influence how parties perceive and respond to each other. This makes it crucial that negotiators adapt quickly to the changing circumstances and new realities with which they are faced.


\end{document}